\documentclass[preprint2]{emulateapj-rtx4}

\slugcomment{To appear in the ApJL.}

\usepackage{txfonts} 
\usepackage{epsfig} 
\usepackage{graphicx}
\usepackage{natbib} 
\bibpunct{(}{)}{;}{a}{}{,}

\begin{document}

\shorttitle{Disks and outflows in the massive protobinary system W3(OH)TW} 
\shortauthors{Zapata, et al.}  
\title{Disks and outflows in the massive protobinary system W3(OH)TW}

\author{Luis A. Zapata\altaffilmark{1}, Carolina Rodr\'\i guez-Garza\altaffilmark{1}, Luis F. Rodr\'\i guez\altaffilmark{1},  
Josep M. Girart\altaffilmark{2}, and Huei-Ru, Chen\altaffilmark{3}}
\altaffiltext{1}{Centro de Radioastronom{\'\i}a y Astrof{\'\i}sica, Universidad 
Nacional Aut\'onoma de Mexico, Morelia 58090, Mexico}
\altaffiltext{2}{Institut de Ciencies de l'Espai (CSIC-IEEC), Campus UAB, Facultat de Ciencies, 
Torre C5-parell 2, 08193 Bellaterra, Catalunya, Spain}
\altaffiltext{3}{Department of Physics, National Tsing Hua University, Hsinchu, Taiwan}

\begin{abstract}
Sensitive and high angular resolution ($\sim$ $0\rlap.{''}7$) (sub)millimeter line and continuum  
observations of the massive star forming region W3(OH) made with the Submillimeter Array
are presented. We report the first detection of two bipolar outflows
emanating from the young and massive "Turner-Welch'' [TW] protobinary system detected 
by the emission of the carbon monoxide. The outflows
are massive (10 M$_\odot$), highly-collimated (10$^\circ$), and seem to be the extended molecular component of 
the strong radio jets and a 22 GHz maser water outflow energized also by the stars in the W3(OH)TW system. 
Observations of the 890 $\mu$m continuum emission and the thermal emission of the 
CH$_3$OH might suggest the presence of two rotating circumstellar disk-like structures associated with the binary system.  
The disks-like structures have sizes of about 1500 AU,  masses of a few M$_\odot$ 
and appear to energize the molecular outflows and radio jets. We estimate that the young stars feeding the outflows 
and that are surrounded by the massive disk-like structures maybe are B-type.  
\end{abstract}


\section{Introduction}

There is recent observational evidence that the formation of massive
stars (early B-type to late O-type) takes place in a way similar to
that of solar-type objects, namely by accretion via a circumstellar
disk: a handful of cases where a disk could be present around a
forming massive star have been presented in the literature
\citep[e.g.][]{fe2011a,fe2011b,gal2010,rod2007,sch2006,pa2005,she1999}.  These
disks are larger than those found in forming solar stars and have
dimensions of less than 1000 AU and masses of a few solar masses.  For
more massive stars (i.e. early O-type), the disks appear to be much more
massive and larger \citep[e.g.][]{zapa2010a,qiu2009,ramon}

One of the nearest and best-studied sites of ongoing massive star
formation is the W3(OH) region. It is located at 2.04 kpc \citep{Hach2006}.  
Within the W3(OH) region there are clearly two objects that host young massive stars, 
W3(OH) itself and the Turner-Welch'' [TW] object, both bright at radio and millimeter 
wavelengths \citep{tu1984}. W3(OH) is a well-known ultracompact
H II region ionized by young OB stars, and rich in OH maser emission 
\citep{reid1995,wil1999,fi2007}. W3(OH)TW or W3(H$_2$O), on the other hand, 
seems to be in a younger state, 
with no associated HII region and with strong dust and molecular emission at 
(sub)millimeter wavelengths \citep{wil1995,wyro1997}.  
\citet{bally1983} reported faint wings in the CO (J=1-0) spectrum
of W3(OH) extending over $\sim$26 km s$^{-1}$. However, no additional
studies for high velocity gas were reported in the literature.

High angular radio observations resolved W3(OH)TW into a
binary system: W3(OH)TW-A and W3(OH)TW-C. The spectrum of W3(OH)TW-A 
from 1.6 to 15 GHz exhibits a power law with spectral index $-$0.6 and the radio emission 
has a large elongation in the east-west orientation suggesting that this 
object is a synchrotron jet \citep{reid1995,wil1999}. The elongated 
continuum source is further coincident, within 0.1 arcsec, with the center of expansion 
of the H$_2$O masers and is aligned with the dominant H$_2$O outflow pattern 
\citep{Alco1993}. At millimeter wavelengths this object shows optically thin dust
emission with a spectral index of $+$3.0 and hot molecular core activity \citep{chen2006,wyro1997}. 
W3(OH)TW-C at radio wavelengths shows a positive spectrum ($+$0.9),
perhaps associated with a thermal radio jet \citep{wil1999}. At millimeter
wavelengths this source is compact and with positive and steep spectrum ($+$3.0) 
related also with optically thin dust emission \citep{chen2006,wyro1997}. Additionally, 
\citet{wyro1997} reported at millimeter wavelengths a third source associated with 
this zone called W3(OH)TW-B.


\begin{figure*}[ht]
\begin{center}
\includegraphics[scale=0.45]{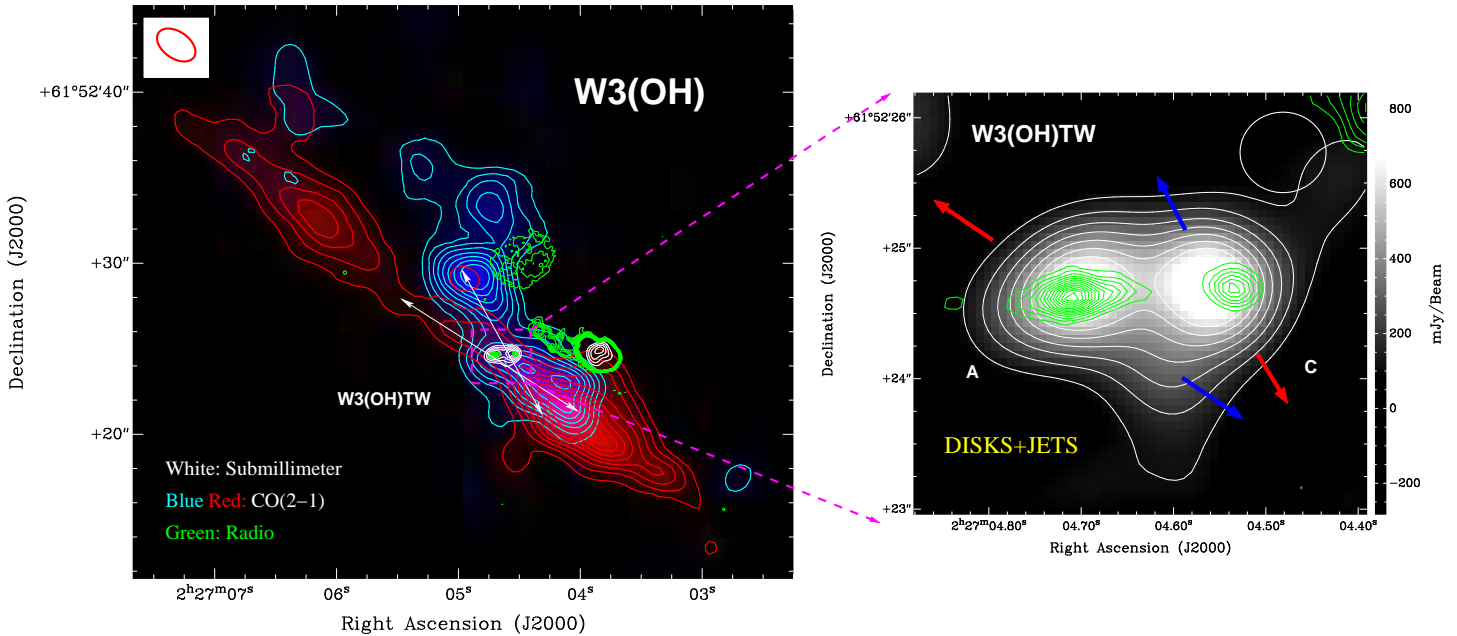}
\caption{\scriptsize LEFT: Integrated intensity color and contour 
  maps of the $^{12}$CO(2-1) emission from the W3(OH) region overlaid
  in contours with the SMA 890 $\mu$m continuum emission (white) and the
  VLA 3.6 cm continuum emission (green).  The blue and red contours are 
  from 30\% to 90\% with steps of 7\% of the peak of the line emission; 
  the peak the $^{12}$CO(2-1) emission is 100 Jy Beam$^{-1}$ km s$^{-1}$. 
  The white contours are from $-$30\% to 90\% with steps of 7\% of
  the peak of the emission; the peak of the 890 $\mu$m map
  is 830 mJy Beam$^{-1}$. The green contours are from 0.45\% to 10\% with steps of 0.1\% 
  of the peak of the emission; the peak at 3.6 cm  
  is 23 mJy Beam$^{-1}$. The synthesized beams of the CO and 890 $\mu$m continuum are described
  in the text and are shown in the left and right upper corners of the two images, respectively. 
  RIGHT: A zoom into the W3(OH)TW region. The green and white
  contours are the same as in the left Figure. The grey scale image 
  shows the SMA 890 $\mu$m continuum emission.  
  The color-scale bar on the right indicates the flux scale in mJy. The blue and red
  arrows represent the position and orientation of the molecular outflows emanating
  from the massive protobinary.}
\label{fig1}
\end{center}
\end{figure*}

\section{Observations}

\subsection{Millimeter}

The observations were made with 8 antennas of the SMA\footnote{The Submillimeter
Array (SMA) is a joint project between the Smithsonian Astrophysical
Observatory and the Academia Sinica Institute of Astronomy and
Astrophysics, and is funded by the Smithsonian Institution and the
Academia Sinica.} on 2007 August in its compact configuration.
The phase reference center for the observations was at R.A. = 02h27m04.30s, decl.=
$+$61$^\circ$52$'$24.5$''$ (J2000.0).
The frequency was centered at 220.730 GHz in the Lower Sideband (LSB), 
while the Upper Sideband (USB) was centered at 230.730 GHz. The primary beam of the 
SMA at around 230 GHz has a FWHM of about 60$''$. The emission from the whole
of the W3(OH) region falls very well inside of our FWHM. 
 
The SMA digital correlator was configured in 24 spectral
windows (``chunks'') of 104 MHz each, with 128 channels distributed
over each spectral window, providing a resolution of 0.812 MHz ($\sim$ 1 km
s$^{-1}$) per channel. 

The zenith opacity ($\tau_{230 GHz}$), measured with the NRAO tipping
radiometer located at the Caltech Submillimeter Observatory, was
from 0.24 to 0.32, indicating reasonable weather conditions during the
observations. Observations of Uranus provided the absolute scale for
the flux density calibration.  The gain calibrators were
the quasars 0102+584 and 0359+509.  The uncertainty in the flux scale
is estimated to be between 15 and 20$\%$, based on the SMA monitoring of quasars.
Further technical descriptions of the SMA and its calibration schemes 
can found in \citet{Hoetal2004}.

The data were calibrated using the IDL superset MIR, originally
developed for the OVRO
\citep{Scovilleetal1993} and adapted for the SMA.\footnote{The MIR-IDL
cookbook by C. Qi can be found at
http://cfa-www.harvard.edu/$\sim$cqi/mircook.html} The calibrated
data were imaged and analyzed in the standard manner using the MIRIAD and KARMA softwares.  
We set the ROBUST parameter of the task INVERT to $-$2 to obtain a slightly better 
resolution sacrificing some sensitivity. The resulting rms noise for the line images was
around 200 mJy beam$^{-1}$ for each velocity channel at an angular resolution 
of $2\rlap.{''}45$ $\times$ $1\rlap.{''}55$ with a P.A. = $65.4^\circ$.

\begin{figure*}[ht]
\begin{center}\bigskip
\includegraphics[scale=0.4]{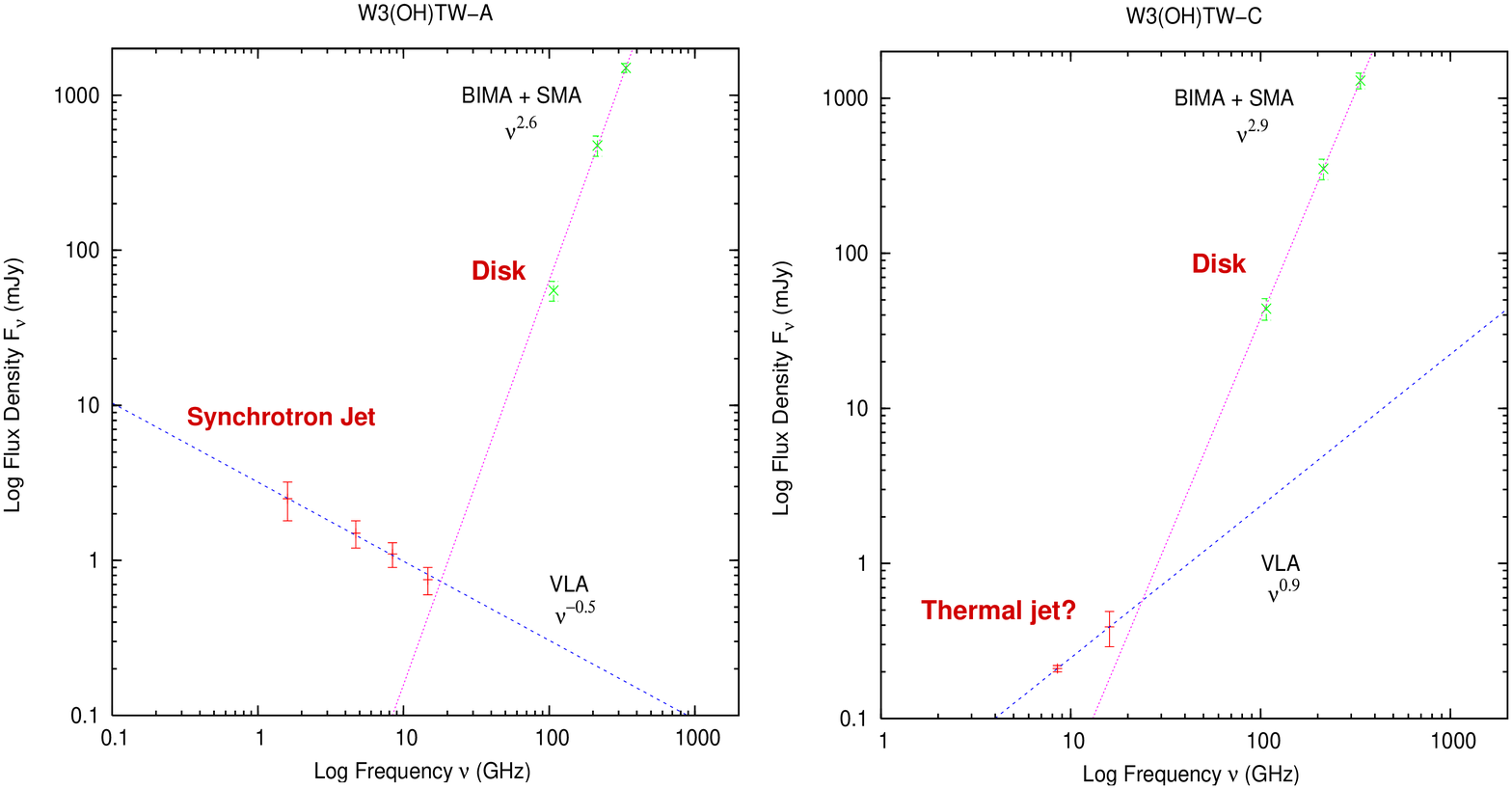}
\caption{\scriptsize Energy spectral distributions for W3(OH)TW-A (left) and W3(OH)TW-C (right) from radio
to (sub)millimeter wavelengths. The radio and millimeter data were obtained from \citet{reid1995}, 
\citet{wil1999}, and \citet{chen2006}. The submillimeter data is presented here (Table 1). 
The line is a least-squares power-law fit (of the form S$_\nu$ $\propto$ $\nu^\alpha$) to the spectrum. 
The $\alpha$-values of the fitting for the different components of the spectrum are shown in the panels.} 
\label{fig1}
\end{center}
\end{figure*}

\subsection{Submillimeter}

The observations were made with 7 antennas of the SMA on 2007 August in its extended configuration.
The receivers were tuned to a frequency of 346.383 GHz in the USB, 
while the LSB was centered on 336.383 GHz. The phase reference center for the observation 
was R.A. = 02h27m03.87s, decl.= $+$61$^\circ$52$'$24.5$''$ (J2000.0).

The zenith opacity ($\tau_{230 GHz}$) was from 0.06 to 0.08,
indicating excellent weather conditions.  Observations of 3C273 with an adopted flux of 8.8 Jy,
provided the absolute scale for the flux density calibration.
The gain calibrators were the quasars 3C84 and 3C111. 
For the continuum, we set the ROBUST parameter to $-$2 to obtain a slightly better 
resolution sacrificing some sensitivity, while for the line, we use ROBUST=$+$2 to obtain
a better signal-to-noise ratio.
The continuum and line images rms noises were around 50 mJy beam$^{-1}$ and
150 mJy beam$^{-1}$, respectively at an angular resolution of $0\rlap.{''}74$ $\times$ $0\rlap.{''}73$ with a
P.A. = $32.7^\circ$ for the continuum and 0.88$''$ $\times$ 0.74$''$ with a P.A. of 80.4$^\circ$ for the line.

\section{RESULTS}

\subsection{Molecular outflows}

In our 2 GHz upper side band of the millimeter observations, we detected the line $^{12}$CO(2-1)
at a rest frequency of 230.53800 GHz. In Figure 1, we present a map of the integrated intensity over
velocity (moment 0) of the line emission observed toward W3(OH). This map was additionally overlaid with the 890 $\mu$m continuum 
emission obtained from these observations and the 3.6 cm radio emission from \citet{reid1995}, and \citet{wil1999}.
The velocity integration is over the velocity ranges: $-$75 to $-$55 km s$^{-1}$ (blue) and $-$45 to $-$25 km s$^{-1}$ (red). 
The emission at ambient velocities ($-$54 to $-$46 km s$^{-1}$) was clearly extended and poorly sampled with the SMA, 
and was suppressed in this moment zero map. 
This map reveals two strong collimated and bipolar outflows emanating from the massive protobinary system W3(OH)TW, both 
with similar orientations. One of the bipolar outflows emanates from W3(OH)TW-A with its blueshifted side
towards the southwest while its redshifted side is located to the northeast. This outflow has a position angle of $+$40$^\circ$. 
The second outflow emanates from W3(OH)TW-C with its blueshifted side towards the northeast while its redshifted side is 
located to the southwest. This outflow has a position angle of $+$15$^\circ$. Both outflows show collimation factors
of about 10$^\circ$. The extension of the outflows is about 0.1 pc.

We note that both red lobes appear to be significantly larger than
the blue lobes. We tentatively suggest that this effect could be created
if very dense molecular material is present between us and the source and
that this gas is stopping the development of the blue lobes.

Assuming that we are in local thermodynamic equilibrium (LTE), the molecular emission is optically thin, an excitation temperature equal 50 K, 
and an abundance ratio of $^{12}$CO/H$_2$ equal to 1 $\times$ 10$^{-4}$, we can estimate the mass of the outflows for the $^{12}$CO molecule 
in the transition $\Delta$J = 2 - 1. We obtained a mass of about 10 M$_\odot$ for each outflow. 

For a mechanical force of F$_M$ = 10 M$_\odot$ 20 km s$^{-1}$/9000 yr = 0.03 M$_\odot$ km s$^{-1}$ yr$^{-1}$ and from the correlation 
presented in \citet{wu2004} for the outflow mechanical force versus the bolometric luminosity of the exciting source, 
we very roughly estimate a luminosity for the central powering source on the order of 10$^{3-4}$ L$_\odot$, which corresponds to 
a massive B-type protostar. This spectral type for the central star is in good agreement with that obtained from the dynamical considerations,
as we will see in the next section.

It is interesting to note that the molecular outflows are very likely be the extended molecular component of the outflows traced 
at smaller scales by the radio jets 
and 22 GHz maser water outflow reported by \citet{reid1995}, \citet{wil1999}, and \citet{Alco1993}. 
However, for example the radio jet associated with the object W3(OH)TW-A has an east-west 
orientation or a  P.A. of $+$90$^\circ$, while the molecular outflow related with this source has a different orientation toward 
the southwest-northeast or a P.A. of $+$40$^\circ$. The 22 GHz maser water outflow has a similar orientation to the radio jet.
This might be explained if the ejected material from W3(OH)TW-A bends sometime after the ejection or maybe this source precesses as expected from
a binary source. There are some cases where the molecular outflow changes orientation or even precesses, see for example 
\citet{cho2006}, \citet{zapa2010b}, and \citet{cunn2009}. Maybe the outflows are arising from different very compact sources within W3(OH)TW-A or 
perhaps this source could be enegizing both outflows as this may precess.

\citet{argon2003} found several OH masers that are associated with the
bipolar outflow traced by the strong H$_2$O masers. These OH masers trace
the outflow at distances of 1-2$"$ from the TW-A source, a factor of 2 larger
that the distances traced by the water masers. Interestingly, the OH masers
suggest that the outflow may be starting to bend in the direction traced
by the CO at scales or 5-10$"$.

\begin{table*}
\begin{center}
\scriptsize
\caption{Physical parameters of the circumstellar disk-like structures \label{tbl-1}}
\begin{tabular}{lccccccc}
\tableline\tableline &\multicolumn{2}{c}{Position$^a$} & Total Flux & & & Disk & Dyn.\\ 
\cline{2-3}             & $\alpha$(J2000) & $\delta$(J2000) & Density (mJy) & Deconvolved Angular Size$^b$ & Spectral & Mass & Mass \\ 
            W3(OH)TW   & 02 27 & $+$61 52  & 0.87 mm & & Index & M$_\odot$ & M$_\odot$\\ 
\tableline 
 A & 04.674 & 24.72  &  1500$\pm$100  & $0\rlap.{''}99 \pm 0\rlap.{''}05 \times 0\rlap.{''}43 \pm 0\rlap.{''}05;~  -70^\circ \pm 5^\circ$& 2.6 & 3.0 & 16\\ 
 C & 04.551 & 24.74  &  1300$\pm$150 & $0\rlap.{''}56 \pm 0\rlap.{''}05 \times 0\rlap.{''}47 \pm 0\rlap.{''}05;~ -90^\circ \pm 10^\circ$ & 2.9 & 2.0 & 6\\ 
\tableline
\end{tabular}
\tablenotetext{a}{Units of right ascension are hours, minutes, and
  seconds and units of declination are degrees, arcminutes, and
  arcseconds.}  \tablenotetext{b}{Major axis $\times$ minor axis;
  position angle of major axis. The values were obtained 
  using the task IMFIT of MIRIAD.} 
\end{center}
\end{table*}

\begin{figure}[ht]
\begin{center}\bigskip
\includegraphics[scale=0.3]{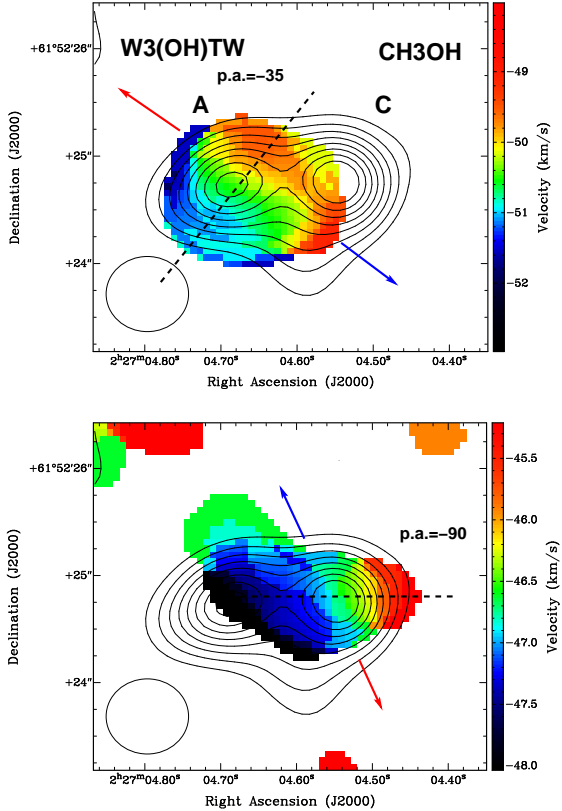}
\caption{\scriptsize UPPER: Integrated intensity weighted velocity map of the 
CH$_3$OH[(3,0)-4(2,2)] emission from W3(OH)TW-A overlaid in contours with the 890 $\mu$m
continuum emission. The integrated velocity range is from 
$-$54 to $-$46 km s$^{-1}$. The systemic velocity of W3(OH)TW-A is about $-$50.5 km s$^{-1}$.
LOWER: Same map as in the upper panel, but now for W3(OH)TW-B. The integrated velocity range is from 
$-$42 to $-$48 km s$^{-1}$. The systemic velocity of W3(OH)TW-B is about $-$46.5 km s$^{-1}$.
The color-scale bars on the right indicates the LSR velocities in km s$^{-1}$. 
The synthesized beam of the line image is shown in the bottom left corner of each image. 
The dashed lines in each panel mark the orientation and position were the PV-diagrams presented in Figure 4
were made. The synthesized beam of the images is 0.88$''$ $\times$ 0.74$''$ with a P.A. of 80.4$^\circ$
and is shown in the left corner of the figures.} 
\label{fig1}
\end{center}
\end{figure}

\subsection{Circumestellar disks?}

\subsubsection{Continuum emission}

In Figure 1, we show the resulting submillimeter ($\lambda$ = 890 $\mu$m) continuum image obtained with
the Submillimeter Array from the high-mass star forming region W3(OH). We detected continuum
emission arising from the ultra-compact HII region W3(OH) itself and the massive protobinary system W3(OH)TW. 
At these wavelengths, the dominant source is W3(OH)TW. The flux density of W3(OH)
is about 1.1 Jy. With our present angular resolution ($\sim$ $0\rlap.{''}7$), we resolve to W3(OH)TW 
into only two strong objects, namely W3(OH)TW-A and W3(OH)TW-C, 
already reported at millimeter wavelengths by \citet[][]{chen2006}, and \citet{wyro1997}. 
However, we do not find strong submillimeter emission (at a level of 4$\sigma$=200 mJy) associated 
with the source W3(OH)TW-B reported by \citet{wyro1997}.
  
W3(OH)TW-A and W3(OH)TW-C are well resolved at these wavelengths 
and show sizes of about 1500 AU at a distance of 2.04 kpc \citep{Hach2006}. 
However, both objects show different morphologies, the objects associated
with W3(OH)TW-A is very elongated in the northeast-southwest direction (with positional angle equal $-$ 70$^\circ$)   
and with a size of its mayor axis of about 2000 AU, while its minor axis has 1000 AU.
W3(OH)TW-C, on the other hand, does not show this marked elongation,
this object instead shows a more roundish morphology, with the size of its minor and mayor axis
quite similar (of about 1000 AU) and with a positional angle equal $-$ 90$^\circ$, see Table 1.

In Figure 2, we present the Spectral Energy Distributions (SEDs) for W3(OH)TW-A and -C.
The spectrum of both objects shows a "combined'' two-regime spectrum, with one component 
observed at radio wavelengths with a negative/positive slowly rising spectrum (with $\alpha$ = $-$0.5 for A 
and $\alpha$ = 0.9 for B) and that are related with  
synchrotron/thermal? jets \citep{reid1995,wil1999}. The second component is
observed at (sub)millimeter wavelengths with the emission that rises rapidly with frequency, 
and is associated with optically thin dust emission from a circumstellar disk 
or maybe an envelope \citep{wyro1997,wyro1999,chen2006}.

We have obtained the spectral indices (S$_\nu$ $\propto$ $\nu^\alpha$) for both structures 
from the millimeter and submillimeter BIMA+SMA observations, see Table 1. 
The spectral indices are very steep (2.6 for the component A and 2.9 for the component B) which 
are consistent with optically thin dust emission  
and with a dust mass opacity coefficient that varies with 
frequency as $\kappa_\nu$ $\propto$ $\nu^{0.6 - 0.9}$. 
This spectral index determination is reliable since 
the 0.89, 1.4, and 2.8 mm observations have similar angular resolution ($\leq$ 1$''$) and
are in very good agreement with those values obtained at millimeter wavelengths by \citet{chen2006}.

With this information, we can estimate the masses of the 890 $\mu$m sources.
Adopting a value of $\kappa_{1.4 mm}$ = 1.5 cm$^2$ g$^{-1}$ (the average of the values of 1.0 cm$^2$ g$^{-1}$, 
valid for grains with thick dust mantles, and 2.0 cm$^2$ g$^{-1}$, valid for grains without mantles). 
This implies  $\kappa_{890 \mu m}$ $\sim$ 2.0 cm$^2$ g$^{-1}$.
Assuming optically thin, isothermal dust emission and a gas-to-dust ratio of 100, and  
a dust temperature of 100 K for the (sub)millimeter objects \citep{chen2006},
we derive masses of about a few solar masses (see Table 1). These values for the mass are in 
reasonable agreement with the values found by \citet{chen2006} for W3(OH)TW-A equal to 5 M$_\odot$
and for W3(OH)TW-C equal to 4 M$_\odot$. The differences would be attributed to the observations of \citet{chen2006}
probably are recovering more extended emission from the objects.

The dimensions ($\sim$ 1500 AU), the masses (of a few M$_\odot$), and that from both sources 
emanate at large scales powerful molecular outflows and at small scales thermal/non-thermal jets 
(Figure 1) suggest that W3(OH)TW-A and -C contain massive circumstellar disks.
Furthermore, the two submillimeter objects have orientations consistent with the ejection
of the outflows (Table 1). 

The derived masses and sizes of the two hypothesized disks (Table
1) will make fairly large optical depths of 0.3 in the dust emission.
In such case, the spectral index may be affected by the optical depth and 
the assumption of optically thin dust emission might be not correct at all. 

There is, however, a problem with the interpretation of W3(OH)TW-A being a massive circumstellar disk. 
The orientation of the synchrotron jet and the H$_2$O maser outflow that emanates 
from this source has an east-west orientation, and is not consistent with the orientation of the disk-like structure. 
However, as mentioned before, maybe W3(OH)TW-A is precessing or the putative disk is actually circumbinary, and 
one compact source in the middle drives the radio jet, and the second one the molecular outflow.

\begin{figure}[ht]
\begin{center}\bigskip
\includegraphics[scale=0.27]{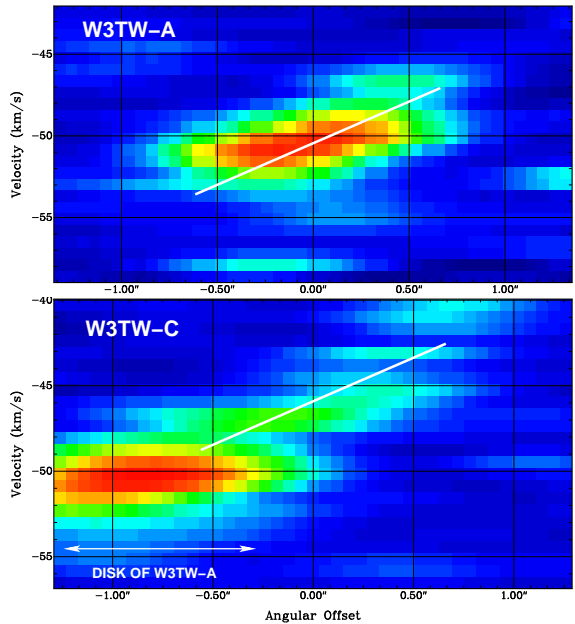}
\caption{\scriptsize Position velocity diagrams computed from 
Figure 3. UPPER: PV diagram for W3(OH)TW-A. LOWER: PV diagram for W3(OH)TW-B. 
The units of the vertical axis are in arcseconds. The synthesized beam of 
the images is 0.88$''$ $\times$ 0.74$''$ with a P.A. of 80.4$^\circ$. 
The spectral resolution is 0.704 km s$^{-1}$. The white lines in the panels
mark the velocity gradient found in every circumestellar disk.} 
\label{fig1}
\end{center}
\end{figure}

\subsubsection{Line emission}

In our 2 GHz lower side band of the submillimeter observations, we detected the line CH$_3$OH[(3, 0)-4(2, 2)] v$_t$=0,1
at a rest frequency of 337.135873 GHz. Figure 3 shows maps of the intensity-weighted velocity 
(moment 1), overlaid with the 890 $\mu$m continuum emission obtained by averaging the line-free channels in our SMA observations.
These maps reveal compact and strong submillimeter molecular emission arising only from the continuum sources and with clear velocity gradients
of a few kilometers per second. W3(OH)TW-A shows a velocity gradient of about 8 km s$^{-1}$ arcsec$^{-1}$ at a position angle of $-$35$^\circ$ and W3(OH)TW-C
of about 6 km s$^{-1}$ arcsec$^{-1}$ at a position angle of $-$90$^\circ$. 
In Figure 3, we only present the strongest CH$_3$OH emission from both sources, however,
in Figure 4 we show more clearly the full magnitude of the velocity gradients. The systemic velocities of each source are a bit different,
 we find that W3(OH)TW-A
is at $-$50.5 km s$^{-1}$, while W3(OH)TW-C is at $-$46.5 km s$^{-1}$. Assuming that the two sources are separated by $\sim1''$ and in a bound circular
orbit, we estimate a lower limit of 30 M$_\odot$ for the whole system. We note that the orientation of the velocity gradient found 
in these objects are in very good agreement with the position of the mayor axis of the circumstellar disk-like structures (Table 1) suggesting 
that we are likely seeing the rotation of the molecular gas in them. 

In Figure 4, we show the kinematics of the molecular gas of both possible circumstellar disks. In the position-velocity diagrams 
it is much more clear to see the gradients and their magnitudes. 
The velocity gradients are clearly linear and associated with the rotation of a rigid body. However,
to detect the Keplerian motions, typically observed in many low-mass circumstellar disks, we need much more angular and spectral resolution.
\citet{zapa2010a} noted that the Keplerian motions of the extremely large circumstellar disk associated with the 
massive young object W51 North are more pronounced close to the protostar while in the edges the molecular gas moves more like a rotating ring. 
Assuming that the velocity gradients are Keplerian, we find that the dynamical mass for W3(OH)TW-A is 16 M$_\odot$ and for W3(OH)TW-C is 6 M$_\odot$. 
The sizes of the line emission are similar to those found in the dust continuum emission.
Those values are in good agreement with \citet{chen2006} that reported a total dynamical mass for the system of about 22 M$_\odot$. However, we note that
they could not separate the velocity gradient of each disk, they instead reported a single gradient across the system.  

Thus, the protostar associated with W3(OH)TW-A has a mass of approximately 13 M$_\odot$, while the associated with
W3(OH)TW-C has a mass of 4 M$_\odot$,  which corresponds to B-type stars. 
 

  

\end{document}